\documentclass[journal]{IEEEtran}
\usepackage{cite}
\usepackage{multirow}
\usepackage{makecell}
\usepackage{color,soul}
\usepackage{gensymb}
\usepackage{dsfont}
\ifCLASSINFOpdf
  \usepackage[pdftex]{graphicx}
  \graphicspath{{../pdf/}{../jpeg/}}
  \DeclareGraphicsExtensions{.pdf,.jpeg,.png}
\else
  \usepackage[dvips]{graphicx}
  \graphicspath{{../eps/}}
  \DeclareGraphicsExtensions{.eps}
\fi
\usepackage[cmex10]{amsmath}
\usepackage{amssymb}

\DeclareMathOperator*{\argminA}{arg\,min}
\usepackage{euscript}
\usepackage{array}

\ifCLASSOPTIONcompsoc
  \usepackage[caption=false,font=normalsize,labelfont=sf,textfont=sf]{subfig}
\else
  \usepackage[caption=false,font=footnotesize]{subfig}
\fi
\usepackage{url}

\begin{document}
%
\title{A Survey on State Estimation Techniques and Challenges in Smart Distribution Systems}
%
%
%

\author{Kaveh~Dehghanpour,~\IEEEmembership{Member,~IEEE,} 
Zhaoyu Wang,~\IEEEmembership{Member,~IEEE,} 
Jianhui Wang,~\IEEEmembership{Senior Member,~IEEE,} 
Yuxuan Yuan,~\IEEEmembership{Student Member,~IEEE,} 
Fankun Bu,~\IEEEmembership{Student Member,~IEEE}
\thanks{This work is supported by the Advanced Grid Modeling Program at the U.S. Department of Energy Office of Electricity under DE-OE0000875.

The authors would like to thank Dr. Ravindra Singh from the Argonne National Laboratory for valuable comments that greatly improved the manuscript.

K. Dehghanpour, Z. Wang, Y. Yuan, and F. Bu are with the Department of
Electrical and Computer Engineering, Iowa State University, Ames,
IA 50011 USA (e-mail: kavehd@iastate.edu; wzy@iastate.edu).

 J. Wang is with the Department of Electrical Engineering, Southern Methodist University, Dallas, TX 75275 USA, and also with the Energy Systems Division, Argonne National Laboratory, Argonne, IL, USA (email: jianhui.wang@ieee.org). 
}}

%
%

\markboth{}%
{Shell \MakeLowercase{\textit{et al.}}: Bare Demo of IEEEtran.cls for Journals}
%

\IEEEpubid{1949-3053 \copyright~2018 IEEE}


\maketitle

\begin{abstract}
This paper presents a review of the literature on State Estimation (SE) in power systems. While covering some works related to SE in transmission systems, the main focus of this paper is Distribution System State Estimation (DSSE). The paper discusses a few critical topics of DSSE, including mathematical problem formulation, application of pseudo-measurements, metering instrument placement, network topology issues, impacts of renewable penetration, and cyber-security. Both conventional and modern data-driven and probabilistic techniques have been reviewed. This paper can provide researchers and utility engineers with insights into the technical achievements, barriers, and future research directions of DSSE.  
\end{abstract}


%
\IEEEpeerreviewmaketitle

\section{Introduction}

\IEEEPARstart{D}ISTRIBUTION System State Estimation (DSSE) is the process of inferring the values of system's state variables using a limited number of measured data at certain locations in the system \cite{Monticelli1999}. Thus, DSSE is basically a numerical process to map data measurements to state variables. While State Estimation (SE) is a well-developed and widely-used concept in transmission systems, its use at the distribution level is still the subject of active research. In recent years we have observed the rapid growth of Advanced Metering Infrastructure (AMI) in electric distribution systems (e.g., according to \cite{FERC2016}, the number of advanced meters in the U.S. was estimated to be 64.7 million devices in 2015, out of a total number of 150.8 million meters, indicating a penetration rate of 42.9\%.) Hence, DSSE is expected to become a significant function in monitoring and power management of smart grids \cite{Primadianto2017}. A general schematic of DSSE function is shown in Fig. \ref{fig:dsse}. Extending conventional SE approaches to active distribution systems is a challenging task due to several factors that are based on the considerable differences between the transmission and distribution systems:
\begin{figure}
\centering
  \includegraphics[width=1\linewidth]{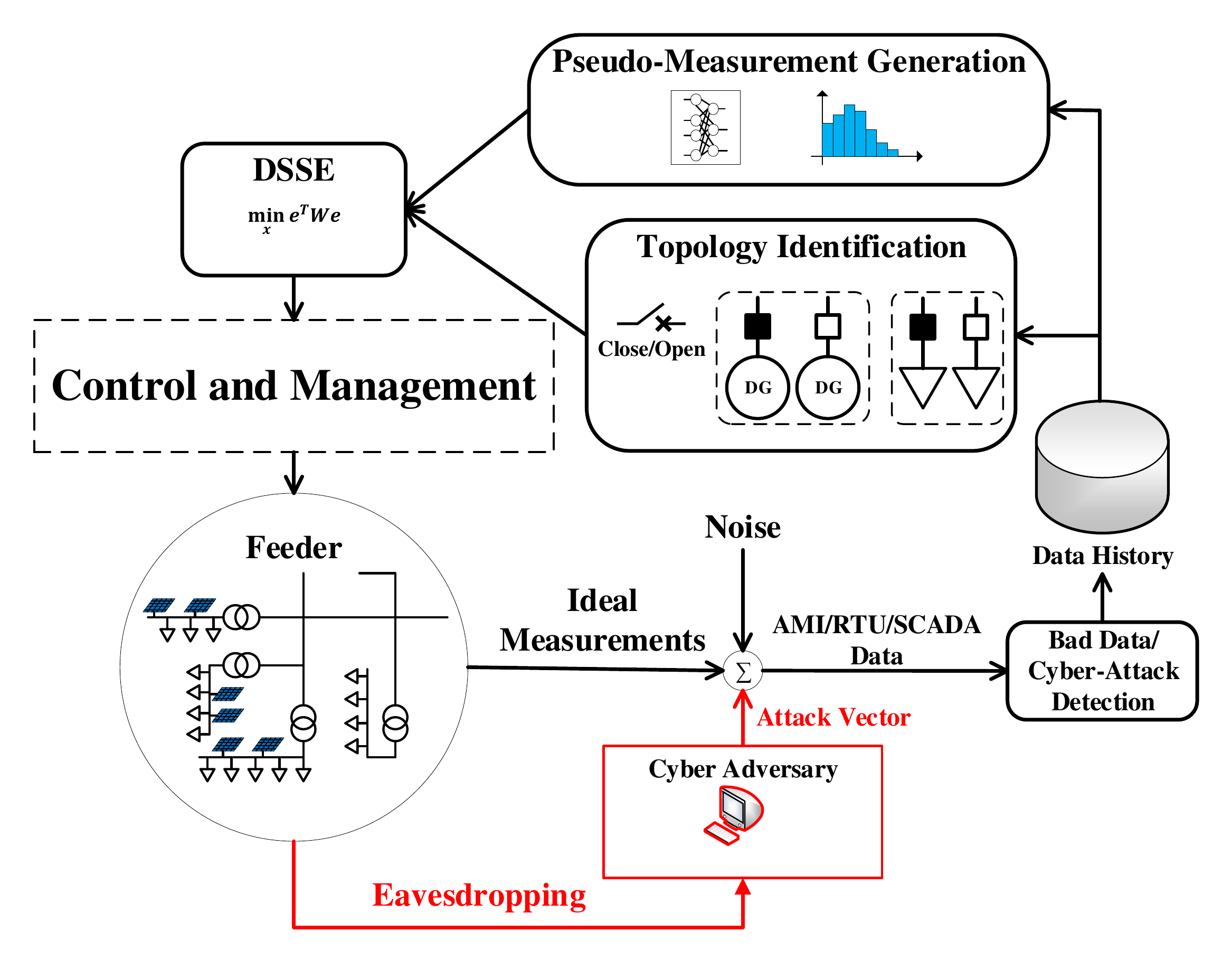}
  \caption{DSSE function in smart grid environment.}
 \label{fig:dsse}
 \vspace{-.5em}  
\end{figure}
\begin{itemize}
\item \textbf{Observability problem:} Unlike transmission systems, the distribution systems are highly unobservable, meaning that the number of metering instruments in a network is generally small compared to the huge size of the system \cite{Bhela2018}.
\item \textbf{Low x/r value:} In distribution systems, we generally face low x/r levels, which render the conventional DC SE techniques in transmission systems unusable at the distribution level \cite{Wu1990}.
\item \textbf{Unbalanced operation:} Distribution systems are in practice highly unbalanced which leads to a higher level of complexity in SE problem formulation. 
\item \textbf{Communication issues:} Constraints on the communication system, such as the network bandwidth and capacity also limit the accuracy and rate of data exchange \cite{Alimardani2015}. 
\item \textbf{Network configuration problem:} Considering the huge size of the distribution network and noting that the complete data related to the topology of this network is not commonly stored an additional degree of complexity to DSSE in these networks \cite{Singh2010}.
\item \textbf{Renewable energy integration:} The higher penetration of renewable power resources introduces a higher level of uncertainty in distribution system operation and DSSE.
\item \textbf{Cyber-security issues:} The issue of cyber-security is a new concern in management and control of active distribution systems.
\end{itemize}

\IEEEpubidadjcol
Despite these challenges, industrial interest in implementing DSSE is growing. Electrical energy firms such as Eaton \cite{Eaton}, Survalent \cite{Survalent}, ETAP \cite{ETAP}, OSI \cite{OSI}, and Nexant \cite{Nexant} have recently devised industrial programs for promoting system monitoring and management at the distribution level for utilities using DSSE. A discussion on relevant experiences on DSSE for radial distribution networks is presented in \cite{Ablakovic}, where the connections between SE implementation and practical variables, such as line lengths, switch flows, voltage regulation, and measurement areas, are elaborated. In this paper we seek to present an extensive review of the proposed solutions to different DSSE-related problems. While the main focus of this paper is DSSE, certain works on transmission system SE have also been cited and reviewed where they become relevant. In summary, this paper discusses the following issues: DSSE problem formulation, pseudo-measurement generation, uncertain network topology, integration of renewable resources, meter placement, and DSSE cyber-security. Special attention has been given to data-driven and machine-learning-based approaches that are gaining interest to address different types of problems \cite{Weng2017-2}.

The reviewed works address critical aspects of DSSE shown in Fig. \ref{fig:dsse}: 1) \textit{DSSE solver module:} in Sections \ref{dsse} and \ref{formul}, we summarize the fundamentals of DSSE, with respect to choice of algorithm and state variables. 2) \textit{Pseudo-measurement generation module:} in Section \ref{observ} the challenge of observability in distribution systems and proposed pseudo-measurement generation solutions in the literature are elaborated. 3) \textit{Topology identification module:} Section \ref{topology} reviews the past works related to online configuration tracking, connectivity detection, and topology discovery, which are pre-requisites for obtaining accurate DSSE solutions. 4) \textit{Feeder and instrumentation module:} The measurement units distributed across the electric power system are the main sources of the information for running the monitoring and control systems. In Section \ref{meter}, the problem of optimal meter placement and potential PMU applications in distribution feeders is presented in terms of practical constraints and objective functions. Modern distribution feeders can have high penetration levels of distributed renewable resources. The impacts of penetration of renewable energy resources in distribution feeders on DSSE are analyzed in Section \ref{renew}. 5) \textit{Cyber-security module:} Reliable DSSE depends on detection and prevention of cyber-intrusions and cyber-attacks. The challenge of cyber-security when performing wide-scale distribution system measurement and monitoring is discussed in Section \ref{cs}. Furthermore, conclusions and future research directions are provided in Sections \ref{conclusion} and \ref{future}. 

\section{Fundamentals of SE}\label{dsse}
\textit{A) Conventional Approach:} Given a measurement vector $\pmb{z}$ (with size $m\times 1$), and a measurement function $\pmb{h}$, which connects the true state vector $\pmb{x}$ (with size $n \times 1$) to the measurement vector (i.e., $\pmb{z} = \pmb{h(x)} + \pmb{e}$, with $\pmb{e}$ denoting the measurement error vector), the state estimation problem can be formulated as a Weighted Least Square (WLS) optimization problem (with bold letters denoting vectors/matrices) \cite{Monticelli1999}:
\begin{equation}
\label{eq:wls}
\pmb{\hat{x}} = \argminA_{\pmb{x}}(\pmb{z}-\pmb{h(x)})^T\pmb{W}(\pmb{z}-\pmb{h(x)})
\end{equation}
where $\pmb{\hat{x}}$ is the estimated state vector, $T$ is the matrix transposition operation, and $\pmb{W}$ denotes the weight matrix that represents the user's confidence in the measured data. A widely-used choice for the weight matrix is $\pmb{W} = diag\{\sigma_1^{-2},...,\sigma_m^{-2}\}$, where $\sigma_j^{2}$ represents the variance of the measurement error corresponding to the $j^{th}$ element of $\pmb{z}$. This choice of the weight matrix is based on two assumptions: 1) the error vector ($\pmb{e}$) has a Gaussian distribution with zero mean, and 2) the measurement errors of different elements of the measurement vector are statistically independent. Under these assumptions the WLS problem transforms to the maximum likelihood estimation. A number of papers have deviated from the conventional approach towards selecting $\pmb{W}$. For instance, in \cite{Chen2016}, using active/reactive power data history, non-diagonal terms have been added to the weight matrix to obtain better WLS accuracy, by modeling the existing correlation between the different measurement samples. This problem has been analyzed in details in \cite{Muscas2014} for modeling the correlations in measurement error distributions of different variables that are measured by the same device (smart meters and PMUs.) For instance, it is shown that the non-diagonal covariance terms between different variables measured by the same device are as follows (notation: active power ($P$), reactive power ($Q$), voltage magnitude ($V$), current magnitude ($I$), power factor ($\cos \Phi$)):
\begin{equation}
\label{eq:covVP}
\begin{split}
&\sigma_{V,P} = \sigma_V^2I\cos\Phi,\ \ \ \sigma_{V,Q} = \sigma_V^2I\sin\Phi\\
&\sigma_{P,Q} = \frac{1}{1}(\sigma_V^2I^2\sin2\Phi-\sigma_\Phi^2I^2V^2\sin2\Phi+\sigma_I^2V^2\sin2\Phi)
\end{split}
\end{equation}

Through another approach, in \cite{Wu2013} and \cite{Hayes2015}, the elements of the diagonal $\pmb{W}$ matrix are updated using a weight function during solution iterations to obtain robustness against bad data. The proposed weight updating mechanism for the $i^{th}$ measurement to obtain new weight value ($\bar{w_i}$) is as follows:
\begin{equation}
\label{eq:weightupdate}
\bar{w_i}=
\begin{cases}
\sigma_i^{-2},\ \ \ \ D'_i\leq k_0\\
\sigma_i^{-2}\zeta_i,\ \ \ \ k_0 < D'_i\leq k_0\\
0,\ \ \ \ D'_i> k_1
\end{cases}
\end{equation}
where, $D'_i$, $\zeta_i$ $k_0$, and $k_1$ are parameters defined based on the residual level corresponding to the $i^th$ data sample. The idea behind (\ref{eq:weightupdate}) is that as $D'_i$ (which is a measure of low quality of the measured data sample) increases beyond the introduced thresholds ($k_0$ and $k_1$), the weight value assigned to it should decrease (with factor $\zeta_i$), reducing the influence of unreliable or bad data samples on the outcome of the WLS.

Conventionally, Gauss-Newton method has been applied to iteratively solve the WLS problem (\ref{eq:wls}) \cite{Wu1990}. This algorithm basically finds a solution to the equation $\nabla J = 0$, where $J$ denotes the objective function of optimization problem (\ref{eq:wls}). The update rules of the algorithm at the $k^{th}$ iteration are as follows:
\begin{equation}
\label{eq:jacobian}
\pmb{H}(\pmb{x}(k)) = \frac{\partial J}{\partial \pmb{x}(k)}
\end{equation}
\begin{equation}
\label{eq:gain}
\pmb{G}(k) = \pmb{H}(\pmb{x}(k))^{T}\pmb{W}\pmb{H}(\pmb{x}(k))
\end{equation}
\begin{equation}
\label{eq:delx}
\Delta \pmb{x}(k) = \pmb{G}(k)^{-1}\pmb{H}(\pmb{x}(k))^{T}\pmb{W}(\pmb{z} - \pmb{h}(\pmb{x}(k)))
\end{equation}
\begin{equation}
\label{eq:update}
\pmb{x}(k+1) = \pmb{x}(k) + \Delta \pmb{x}(k)
\end{equation}
where, $\pmb{H}$ is the Jacobian of $J$ with respect to the state variables, and $\pmb{G}$ is the system gain matrix. Other algorithms, such as back tracking method, trust region method, and quasi-Newton techniques, have also been applied instead of the classical Gauss-Newton method, to obtain better convergence properties \cite{Nie2016}. Noting the non-convexity of (\ref{eq:wls}) and the sensitivity of Newton method to initial conditions and gain matrix ill-conditioning, in \cite{Weng2012} and \cite{Yao2}, a Semi-Definite Programming (SDP) approach is proposed to find a good initial guess for the Newton method. The SDP formulation is based on the convex relaxation of the original WLS problem, which also guarantees the existence of a unique global solution. The computational efficiency of SDP is shown to be superior compared to that of the original non-convex problem. To further improve the computational performance of SDP-based SE, distributed algorithms have been employed for obtaining a solution \cite{Zhu2014}.

Another modification in the structure of WLS (\ref{eq:wls}) is the inclusion of \textit{virtual measurements} as equality constraints ($\pmb{c(x)} = \pmb{0}$). Virtual measurements represent operator's perfect information on certain aspects of system operation (e.g., zero-power-injection at nodes without customers.) Lagrange multipliers ($\pmb{\lambda}$) have been proposed as penalty factors for enforcing these equality constraints \cite{Lin1996}. The modified WLS objective function is defined as follows:
\begin{equation}
\label{eq:wlsLag}
\{\pmb{\hat{x},\hat{\lambda}}\} = \argminA_{\pmb{x},\pmb{\lambda}}(\pmb{z}-\pmb{h(x)})^T\pmb{W}(\pmb{z}-\pmb{h(x)}) + \pmb{\lambda}^T\pmb{c(x)}
\end{equation}

Given the above objective function, the state update step in the Gauss-Newton method (\ref{eq:delx}) is changed to:
\begin{equation}
\label{eq:delxLam}
\begin{bmatrix}
\Delta \pmb{x}(k)\\
\pmb{\lambda}(k)
\end{bmatrix} = 
\begin{bmatrix}
\pmb{H}^T\pmb{W}\pmb{H}&\pmb{C}(\pmb{x}(k))^T\\
\pmb{C}(\pmb{x}(k))&0
\end{bmatrix}^{-1}
\begin{bmatrix}
\pmb{H}^T\pmb{W}(\pmb{z} - \pmb{h}(\pmb{x}(k)))\\
-\pmb{c}(\pmb{x}(k))
\end{bmatrix}
\end{equation}
where, $\pmb{C(x)} = \frac{\partial\pmb{c(x)}}{\partial\pmb{x}}$.

\textit{B) Alternative DSSE Structures:} While WLS represents the conventional SE in power systems, alternative mathematical formulations have been proposed for the purpose of increasing the robustness of the state estimator when facing bad data. Noting the susceptibility of WLS to bad data, in \cite{Mili1994}, the use of Least Median of Squares (LMS) and Least Trimmed Squares (LTS) is studied, which shows improved behavior in handling outliers. Also, \cite{Gol2014} investigates the use of Least Absolute Value (LAV) estimator, which has the property of automatic bad data rejection. Increasing the robustness of SE has also been promoted by using a Generalized Maximum-likelihood (GM) estimator instead of WLS in \cite{Zhao2017}, where normalized residuals ($r_n$) are used through a convex score functions (denoted as $\rho(.)$) in formulating the objective function. The SE formulation for these different approaches (including pros and cons) are shown in Table \ref{table:algs}, in terms of the objective function in optimization problem (\ref{eq:wls}). In this table, the residuals $\pmb{r} = [r_1,...,r_m]^T$ are defined as $r_i = z_i - h_i(\pmb{x})$. Also, $med\{\}$ and $r_{(i)}$ define the set median and the $i^{th}$ order statistics, respectively. Numerical comparisons of these alternative DSSE formulations in terms of robustness against system parameter uncertainties are presented in \cite{Kuhar}.  
\begin{table}[h]
\begin{center}
 \caption{Available Robust SE Formulations}
    \begin{tabular}{ |m{0.35in}|m{1in}|m{1.5in}|}
    \hline
    \textbf{Method} & \textbf{Objective Function} & \textbf{Pros and Cons}\\ \hline
    WLS&$\pmb{r}^T\pmb{W}\pmb{r}$&(+) Fast, simple, widely-used,\ \ \ \ \ \ (-) Sensitive to bad data\\ \hline
 LMS&$med \{r_1^2,...,r_m^2\}$&(+) Robust against bad data and leverage points,\ \ \ \ \ \ \ \ \ \ \ \ \ \ \ \ \ \ \ \ \ \ \ \ \ \ \ \ \ \ \ \ \ (-) High computational cost, high measurement redundancy requirements\\ \hline
    LTS&$\sum_{i=1}^{h}r_{(i)}^2$&(+) Robust against bad data,\ \ \ \ \ \ \ \ \ \ \ \ \ \ \ \ \ \ (-) High computational cost and memory requirement\\ \hline
    LAV&$\sum_{i=1}^{m}|r_i|$&(+) Robust against bad data, small sensitivity to line impedance uncertainty,\ \ \ \ \ \ \ \ \ \ \ \ \ \ \ \ \ \ \ \ \ \ \ \ \ \ \ \ \ \ \ \ \ \ \ \ \ \ \ \ \ \ \ \ \ \ \ \ \ \ \ \ \ \ \ \ \ \ \ \ (-) High computational cost, sensitivity to leverage points and measurement uncertainty\\ \hline
    GM&$\sum_{i=1}^{m}\sigma_i^{-2}\rho(r_{n_i})$&(+) Robust against bad data,\ \ \ \ \ \ \ \ \ (-) Parameter selection sensitivity\\ \hline
    \end{tabular}
    \label{table:algs}
\end{center}
 \vspace{-1.25em}
\end{table}

Other approaches towards structuring the DSSE have been presented as well. For instance, some works in the literature tend to propose estimators which relax the Gaussian uncertainty assumption inherent to WLS. This is of practical importance given that this assumption is shown, through field tests, to be largely inaccurate \cite{Shaobu}. Using Mean Squared Estimator (MSE) an analytic SE formulation is obtained in \cite{Bilil2018} which does not depend on Gaussian uncertainty assumptions and is capable of bad data measurement detection. A similar estimator is used in \cite{Angioni2016}, where a Bayesian alternative to WLS is proposed. It is shown that the Bayesian approach has specifically better performance in presence of non-Gaussian uncertainty. Unlike WLS (equation (\ref{eq:wls})), the Bayesian approach tends to estimate states as a conditional averaging operation: 
\begin{equation}
\label{eq:bayes}
\pmb{\hat{x}} = E\{\pmb{x}|\pmb{z}\} = \int \pmb{\alpha} f_{\pmb{\alpha}|\pmb{z}}(\pmb{\alpha}|\pmb{z})d\pmb{\alpha}
\end{equation}

Calculating $E\{\pmb{x}|\pmb{z}\}$ depends on our knowledge of the distribution function $f_{\pmb{x}|\pmb{z}}$, which can be obtained using Bayes rule, the measurement functions, and statistical properties of the system. Citing availability of accurate knowledge of second order statistics as a shortcoming of MSE-based methods, in \cite{Genes2018} an alternative DSSE formulation is presented as a \textit{matrix completion} problem which can be efficiently solved for billions of entries. Using information-theoretic reasoning it is shown that the optimal performance of DSSE is bounded by the capacity of AMI communication channels in charge of transmitting measurement samples to system operator.     

To reduce the size of the optimization problem and speed up the convergence of WLS for large-scale feeders, in \cite{Jesus2015}, the concept of quasi-symmetric impedance matrix is employed. This is achieved by adding the following constraint to the conventional WLS: 
\begin{equation}
\label{eq:trx}
\begin{split}
&\min_{\pmb{x}} (\pmb{z}-\pmb{h(x)})^T\pmb{W}(\pmb{z}-\pmb{h(x)})\\
&s.t.\ \pmb{g_0(x)} = \pmb{x} - \pmb{x_0} - \pmb{TRX}\cdot\pmb{I(x)} = 0\\
\end{split}
\end{equation}
where, $\pmb{x}$ and $\pmb{x_0}$ represent the voltage node state vector and the substation voltage, respectively. $TRX$ denotes the reduced impedance matrix and $\pmb{I}$ is the set of nodal current injections.

\section{DSSE Problem Formulation}\label{formul}
Due to the basic differences between transmission and distribution systems, the DSSE problem formulation can have major deviations from the conventional SE. The main point of difference is the modeling of measurement function ($\pmb{h}$) in DSSE, as this function reflects the power flow equations in the power system. Hence, based on the choice of state and measured variables, choice of AC versus DC Power Flow (PF), and the representation of phases in power flow equations (for application in unbalanced systems), the measurement function can have different forms. In this section, we review the two basic formulations of DSSE (in terms of choice of state variables and measurement function) provided in the literature.

\textit{A) Voltage-Based DSSE:} Traditionally, bus voltage magnitude and phase angle values have been used as state variables in transmission systems \cite{Monticelli1999}. This conventional approach has also been employed in DSSE \cite{Baran1994} \cite{Lu1995} \cite{Haughton2013} \cite{Deng2002}.

\begin{table*}
\begin{center}
 \caption{Available DSSE Formulation Structures}
  \includegraphics[width=0.7\linewidth]{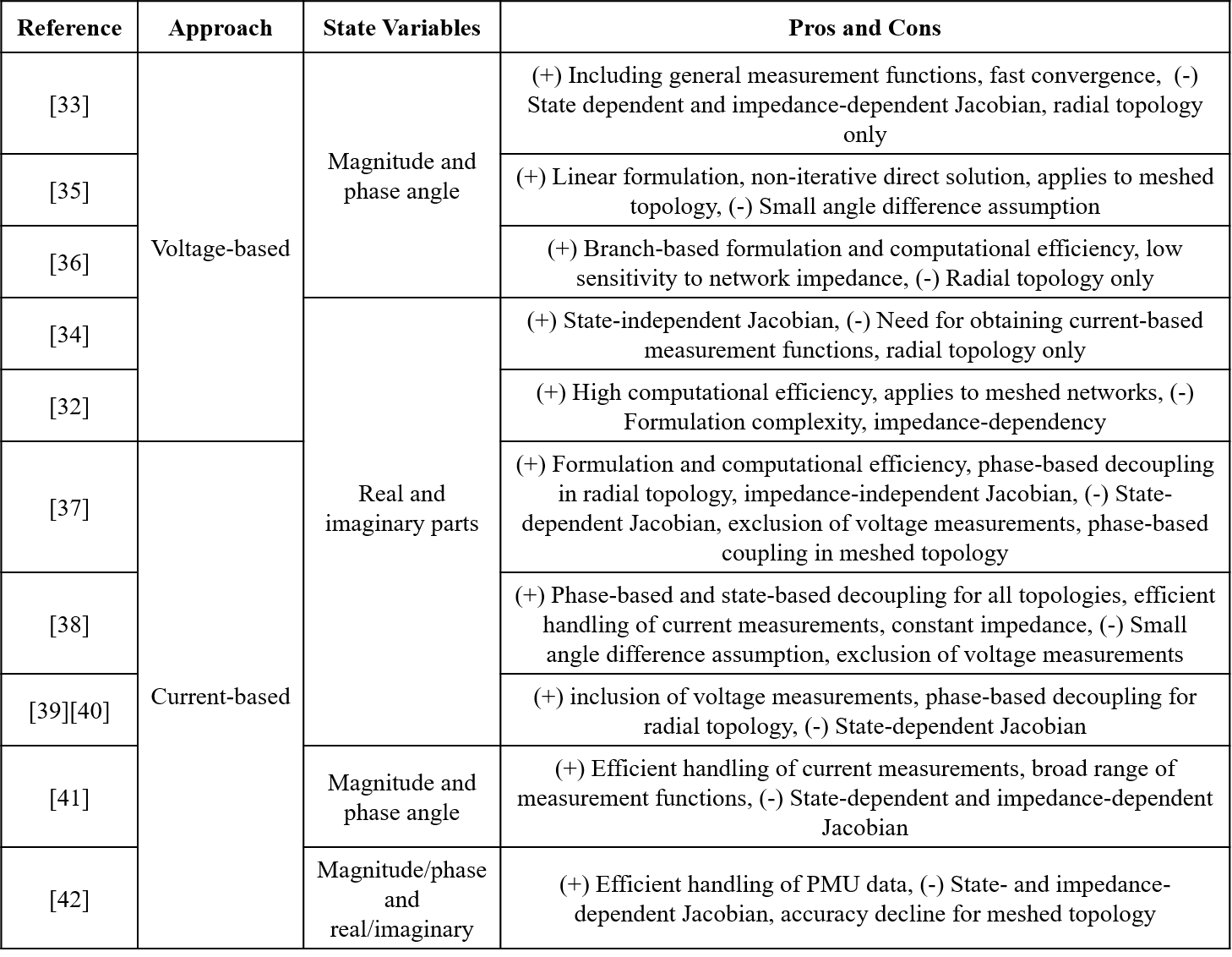}
    \label{table:form}
\end{center}
 \vspace{-1.25em}
\end{table*}

\textit{B) Branch-Current-Based SE (BCSE):} A notable group of works, have adopted branch current as state variables, which turns out to be a more natural way of DSSE formulation for distribution systems \cite{Baran1995} \cite{Lin2001} \cite{Teng2002} \cite{Baran2009} \cite{Wang2004} \cite{Pau2013}. A summary of the properties of different DSSE formulations is shown in Table \ref{table:form}.

\section{Distribution System Observability}\label{observ}
``Observability'' refers to the system operator's ability to solve the state estimation problem. This depends on the number and location of metering instruments in the power system. Also, the availability and quality of critical measurement data samples in real-time has a crucial impact on power system observability. Conventionally, numerical and topological methods have been used to assess the observability of transmission systems with respect to the number and location of meters, as demonstrated in \cite{Monticelli1999}. Alternative observability assessment procedures have been employed at distribution level. For instance, in \cite{Brinkmann2017} a probabilistic approach is adopted to define an Unobservability Index (UI) as follows:
\begin{equation}
\label{eq:UI}
UI = \sum_{i=1}^{n} K_i = \sum_{i=1}^{n} (\sum_{j=1}^{B_i}-p(b_{i,j})\log_2 p(b_{i,j}))
\end{equation}
where, $K_i$ denotes the entropy of the $i^{th}$ state (with $p(b_{i,j})$ defining the probability of the $j^{th}$ bin for the $i^{th}$ state.) Basically, UI represents our overall uncertainty on the distribution system state variable values. As another example, a graph-theoretic criterion for local observability assessment of distribution networks is obtained in \cite{Bhela2018}.

Unlike transmission systems that enjoy a high level of data redundancy, the distribution systems are generally under-determined with poor observability. Thus, the accuracy of DSSE can be highly affected by the quality and availability of sensor data. The distribution system can easily become unobservable in case of communication failure/delays. Hence, bad/missing measurement data is closely connected to measurement redundancy and preserving the reliability of the DSSE problem. ``Bad'' data refer to data measurements that have considerable deviation from the underlying actual behavior, due to meter malfunction and communication noise. Missing data can also be treated as a special case of bad data. Conventionally, at the transmission level, bad data detection has been performed by inspecting the normalized measurement residuals. However, this method is subject to failure and complications in case of insufficient measurement redundancy and multiple sources of bad data \cite{Monticelli1999}. Hence, alternative approaches have been employed to address this problem, along with the sub-problem of missing data, at the distribution level (refer to Section \ref{dsse}.)

Hence, to improve the observability of distribution systems, the input measurement set needs to be artificially augmented (to compensate for missing data) or corrected (to compensate for bad data.) This can be done through employing ``pseudo-measurement'' samples, which are artificially-generated data-points (e.g., active/reactive power, voltage and current, etc.) based on the data history of the distribution systems \cite{Wu1990}. A basic approach is to use standard load profiles for generating pseudo-measurements \cite{Angioni2016-2}. Given that these data-points are not highly accurate, they introduce high variance levels in the weight matrix ($\pmb{W}$), which could even lead to ill-conditioning of the DSSE problem. Data-driven approaches are employed for generating pseudo-measurements and handling their uncertainty, including probabilistic and statistical analysis, and machine-learning-based techniques.

\begin{table*}
\begin{center}
 \caption{Available literature on pseudo-measurement generation}
  \includegraphics[width=0.7\linewidth]{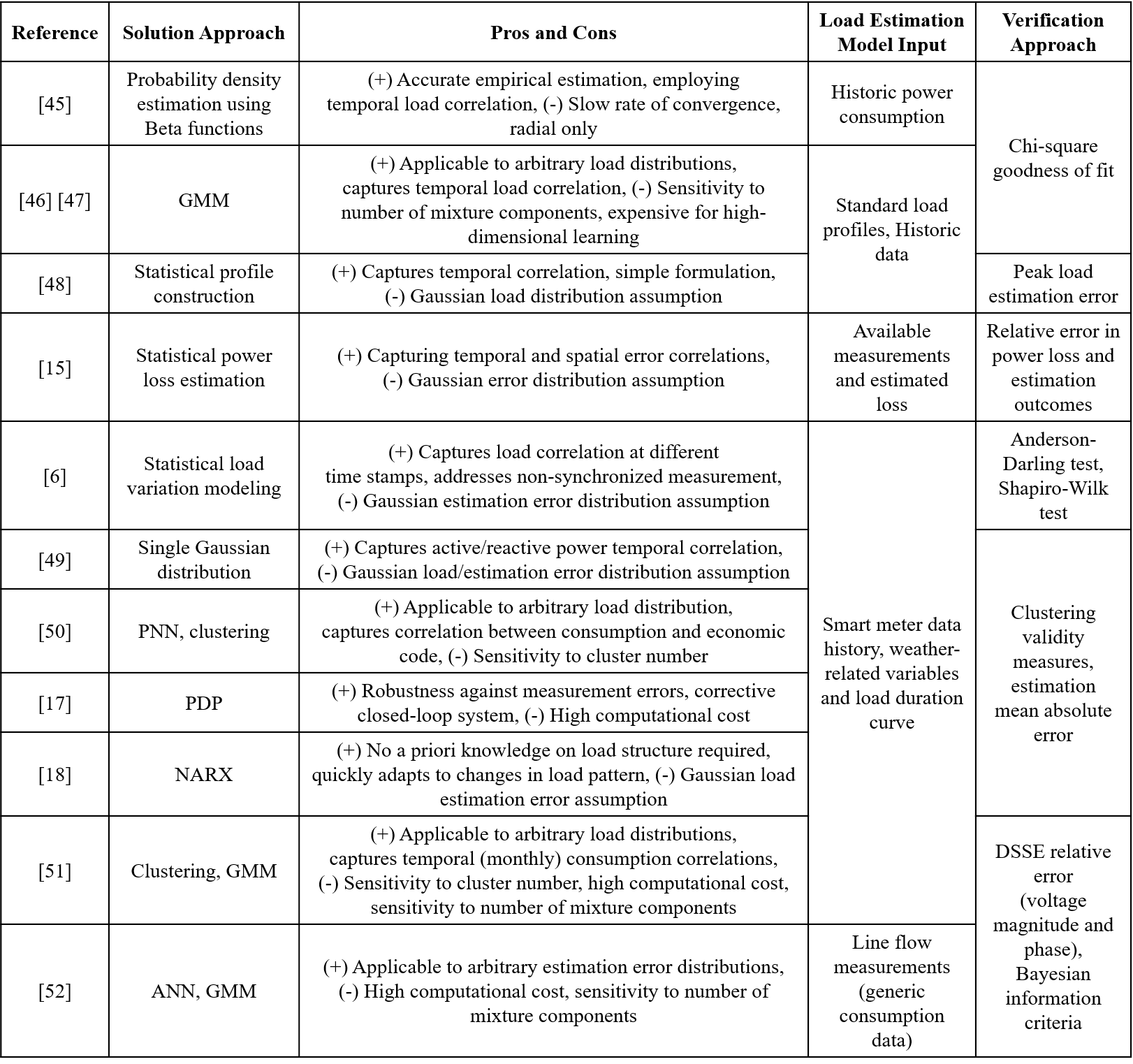}
    \label{table:PM}
\end{center}
 \vspace{-1.25em}
\end{table*}

\textit{A) Probabilistic and Statistical Approaches:} Methods based on probabilistic and statistical techniques, which employ spatial/temporal correlation and historic probability distribution data, are widely used for generating reasonable pseudo-measurements and assessing their uncertainty. This includes empirical studies \cite{Ghosh1997}, Gaussian Mixture Models (GMMs) and Expectation Maximization (EM) \cite{Singh2010-2} \cite{Singh2009}, time-varying variance and mean modeling \cite{Angioni2016-2}, correlation analysis (between total and individual consumption) \cite{Nguyen2015}, nodal active-reactive correlation analysis \cite{Chen2016}, internodal and intranodal correlation modeling \cite{Muscas2014}, intertemporal correlation analysis \cite{Alimardani2015}, multi-variate complex Gaussian modeling \cite{Arefi2015}, and constrained optimization \cite{Dzafic2017}.

\textit{B) Learning-Based Approaches:} Machine learning algorithms have also attracted scientific attention in solving DSSE problems, including addressing the problem of active/reactive power pseudo-measurement generation and uncertainty assessment. Probabilistic Neural Networks (PNNs) \cite{Gerbec2005}, Artificial Neural Network (ANN) \cite{Manitsas2012}, clustering algorithms \cite{Gahrooei2018}, Parallel Distributed Processing networks (PDP) \cite{Wu2013}, and Nonlinear Auto-Regressive eXogenous (NARX) \cite{Hayes2015}. 

A summary of the notable papers in these two categories are shown in Table \ref{table:PM}. Pseudo-measurement generation is basically a special type of load estimation at distribution level. While there is a considerable number of works done in this area, still unanswered questions remain. For instance, most of the papers, instead of using real AMI data history, rely on standard load profiles to perform numerical analysis and verification. Also, the huge amount of data in practice can cause certain learning methods to become computationally expensive. Managing this ``big data'' challenge in distribution systems requires further research and studies.

\section{Network Topology and Configuration}\label{topology}
\begin{figure}
\centering
  \includegraphics[width=0.8\linewidth]{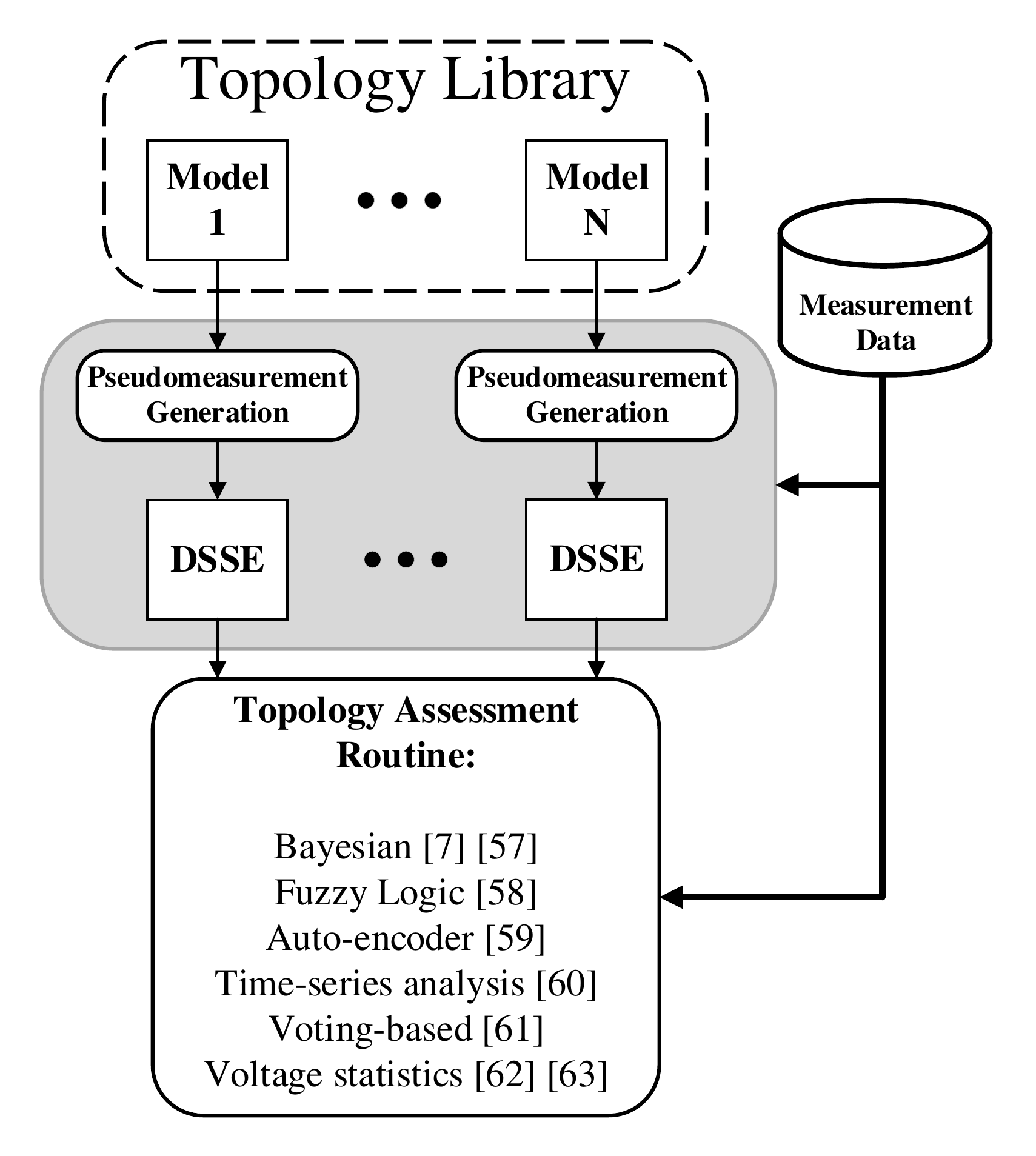}
  \caption{Data-driven system configuration detection.}
 \label{fig:topology}
 \vspace{-.5em}  
\end{figure}

The topology identification problem can be categorized into two separate, yet related, subproblems: 

\textit{A) System configuration identification:} The basic assumption within this set of problems is that the basic topology of the network is known to the system operator. However, due to local events (such as faults, line disconnections, switching events, etc.) the basic topology will undergo local changes over time. Limited knowledge of the operator on these changes will affect the accuracy of SE solutions. Hence, the objective is to use the system-wide measurements to update our knowledge of system configuration to avoid topology errors (i.e., state of switches, fuses, lines, DG/customer connection status.) Conventionally, generalized SE models have been used at the transmission level (with switch-related variables added to the SE formulation) to detect and correct topological errors \cite{Monticelli1999} \cite{Lourenco2015}. Similar classic methods have been applied to DSSE as well \cite{Korres2012} \cite{Baran2009-2}. Apart from the classical approaches, other probabilistic and data-driven methods have been applied for topology detection and identification in distribution systems. These methods are usually based on a data-driven search process in a limited topology space (i.e., topology library) defined by variations on the basic topology, as shown in Fig. \ref{fig:topology}. Probabilistic recursive Bayesian approach \cite{Singh2010} \cite{Hayes2016}, fuzzy-based pattern recognition \cite{Singh2005}, auto-encoders \cite{Miranda2012}, PMU voltage time-series \cite{Cavraro2017}, voting technique (``vote'' for the best candidate structure) \cite{Arghandeh2015}, correlation analysis \cite{Luan2015}, and maximum likelihood estimation \cite{Cavraro2017-2}, are a few of the proposed topology search methods. 

\textit{B) Topology learning:} Another set of problems are based on the assumption that the system operator has very limited or no knowledge of the basic topology of the network (which is highly applicable to the secondary distribution networks.) The objective is to discover the topology of the network by relying on nodal and branch measurements. Graph-theoretic algorithms have been used widely for topology discovery and learning considering different assumptions on system operator's knowledge on topology. A sparse graph recovery model has been adopted in \cite{Babakmehr2016} to perform topology discovery, based on DC PF. The proposed method, which is based on nodal measurements, requires no \textit{a priori} information on the topology of the network. Another data-driven graphical approach towards topology learning is proposed in \cite{Weng2017}. In this work, an efficient graphical model is developed to represent the voltage magnitude dependencies (using mutual information as a measure of affinity) between neighboring buses (the basic assumption in this work is that current injections are statistically independent.) This method only depends on statistics of nodal voltage magnitude measurements (smart meter data) to reconstruct the partially or fully unknown radial or weakly-meshed topology. It is shown that for a radial feeder, the spanning tree that maximizes measures of internodal voltage mutual information corresponds to the true topology of the system. In \cite{Deka2017}, using nodal voltage measurements, the authors have been able to learn the topology of a radial feeder using mutual statistical properties of the measured variables. The proposed model is based on a linear approximation of lossless AC PF, and employs a bottom-to-top approach, in which the structure learning begins with the end nodes and moves towards the substation by choosing the proper parent nodes at each stage. The method is shown to have acceptable performance under a wide variety of assumptions, including no prior knowledge on the basic topology and missing measurement data. In \cite{Jayadev2017}, graph-theoretic interpretation of principal component analysis and energy conservation are employed in the context of graph theory to obtain radial distribution system topology through smart meter energy usage data. A more general approach (applicable to meshed networks even with missing PMU phase measurements) for estimating both the topology of the network and the line parameters is proposed in \cite{Yu2017}, where the line parameters and system topology are updated consecutively through an EM-based approach. Starting with an initial topology guess, at each step of the algorithm, the topology is updated by removing edges with small estimated susceptance values to improve the estimation likelihood.

\section{Distribution Network Metering System Design and Analysis}\label{meter}
\subsection{Metering Instrument Placement}\label{place}

\begin{table}
\begin{center}
 \caption{Meter Placement Methods}
  \includegraphics[width=1\linewidth]{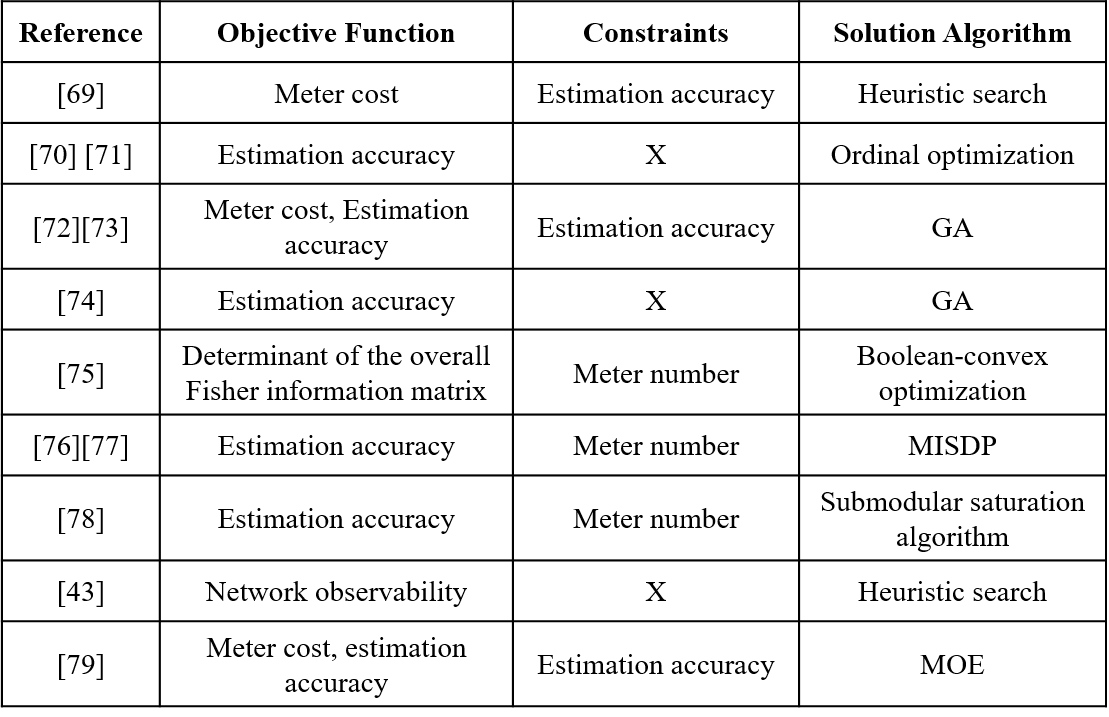}
    \label{table:meter}
\end{center}
 \vspace{-1.25em}
\end{table}
Optimizing the location of metering instruments in distribution systems is a significant subject for research, given the size of the system and potentially limited financial resources \cite{Baran1996}. Different objectives have been proposed in the literature to address this problem, including improving system observability, minimizing installation/maintenance costs, bad data detection capability, and improving the DSSE accuracy \cite{Liu2002} \cite{Wang2004} \cite{Singh2009-2} \cite{Singh2011} \cite{Liu2012} \cite{Celli2014} \cite{Kumar2005} \cite{Xygkis2017} \cite{Xygkis2017-2} \cite{Damavandi2015} \cite{Brinkmann2017} \cite{Prasad2018} \cite{Yao}. Different algorithms have been tried for solving the placement problem, including Genetic-Algorithm (GA), Mixed Integer Linear Programming (MILP), Mixed Integer Semi-Definite Programming (MISDP), and Multi-Objective Evolutionary (MOE) methods. A summary of the different meter placement approaches is given in Table \ref{table:meter}. 

\subsection{PMU Applications and Impacts on DSSE}\label{pmu}
PMUs are able to provide synchronized voltage, power, and current measurements that enable accurate tracking of state variables and efficient control and management decisions. Also, generally the sampling frequency of PMUs (up to 30 kHz) is much higher than that of smart meters (0.277 mHz - 16.7 mHz), which leads to system observability on a higher temporal granularity. However, compared to smart meters, the use of PMUs in distribution networks is still very restricted due to their prohibitive costs. Hence, a critical research direction related to PMUs is optimizing the number and location of PMUs to enhance system observability, while limiting the measurement infrastructure costs \cite{Carta} (also see Section \ref{place}).

In terms of application in distribution systems, PMUs have been employed for high-resolution voltage/power profiling, oscillation detection, topology identification, and event detection, as outlined in \cite{Meier2017}. On the other hand, smart meters have been used mostly for low-resolution load forecasting and management, and connection verification \cite{Wang}. In terms of algorithm design for DSSE and topology identification, one considerable difference between the methods proposed for systems with only smart meters and systems with PMUs is the ``small phase angle difference assumption''. Hence, due to unavailability of phase angle data in absence of PMUs many papers have assumed that the nodal voltage phase angles in a system are almost equal \cite{Weng2017,Deka2017,Peppanen2016}. While this assumption introduces bounded inaccuracies in the final estimation/identification outcomes, it enables system operators to monitor the state of distribution systems without PMUs. Furthermore, adding the voltage phase data or flow measurements can highly improve the estimation and identification routines' performance.

\section{Penetration of Renewable Resources}\label{renew}
A few papers have analyzed DSSE under high penetration rates of renewable power. The main source of challenge in performing SE in presence of renewable resources is their uncertain output power \cite{amini2018trading}. Also, deep penetration of renewable power sources affect the voltage profile of distribution systems. This stresses the need for more advanced voltage monitoring capabilities \cite{Dzafic2017}. In case of pseudo-measurement generation for these resources, it is believed that the non-Gaussian distribution of renewable power would adversely affect conventional WLS-based DSSE methods. Moreover, as shown in \cite{Weng2017-2}, fast changes in system state can result in the WLS-based DSSE to get trapped in local minima with errors as high as $10^5$ times the underlying global solution. Also, given that the performance of conventional Gauss-Newton algorithm highly depends on the initial conditions, finding good initial conditions for DSSE in systems with deep renewable penetration is a difficult task \cite{Bhela2018}. To address these challenges several papers have adopted different approaches for solving the SE (in general) and DSSE (in particular) in presence of renewable-based DGs.

Probabilistic methods represent the major group of techniques for modeling the impacts of renewable uncertainty on SE. A forecasting-aided SE mechanism is proposed in \cite{Zhao2016} to capture the temporal and spatial correlation among DGs and loads for their short-term prediction (to be used as pseudo-measurements in SE), using a linear autoregressive model. In \cite{Bilil2018}, another forecasting-aided SE method is proposed to manage the uncertainties of load and renewable resources based on a GMM technique for obtaining the non-Gaussian distribution of renewable power while incorporating the dynamics of the system. Moreover, this estimator shows good performance even with limited data, which makes it a promising candidate for DSSE. As an extension to \cite{Liu2012}, the effect of the uncertainty of renewable DG power profile on meter placement has been modeled in \cite{Liu2014} using GMM. A probabilistic graphical modeling technique has been proposed in \cite{Weng2017-3} for capturing short term uncertainty of SE in systems with high PV penetration. The physical governing laws of the system (i.e., PF equations) have been embedded into the SE model. A distributed belief propagation method is performed for state inference, which yields superior results compared to the conventional deterministic WLS method. Another probabilistic approach is adopted in \cite{Angioni2016-2} for pseudo-measurement generation in networks with high residential PV penetration using Beta distribution functions. It is speculated that the uncertainty of PV systems has the highest impact on the DSSE at mid-day time intervals (when usually the load profile is not peaking.) To model the non-Gaussian uncertainty of PV power in DSSE, pseudo-measurements are generated (with 15-minute time resolution) for roof-top dispersed PV systems employing a weather-dependent model for constructing general PV power probability density functions, considering solar radiation, temperature, number of arrays and their physical characteristics. This approach shows considerable improvements on DSSE accuracy compared to using conventional standard profiles. While in \cite{Angioni2016-2} the possible correlation between physically nearby renewable DGs are not modeled, it is demonstrated in \cite{Muscas2014} that including the correlation between close DGs for pseudo-measurement generation leads to further improvements in DSSE accuracy.
\begin{table}
\begin{center}
 \caption{Available literature on SE cyber-security}
  \includegraphics[width=1\linewidth]{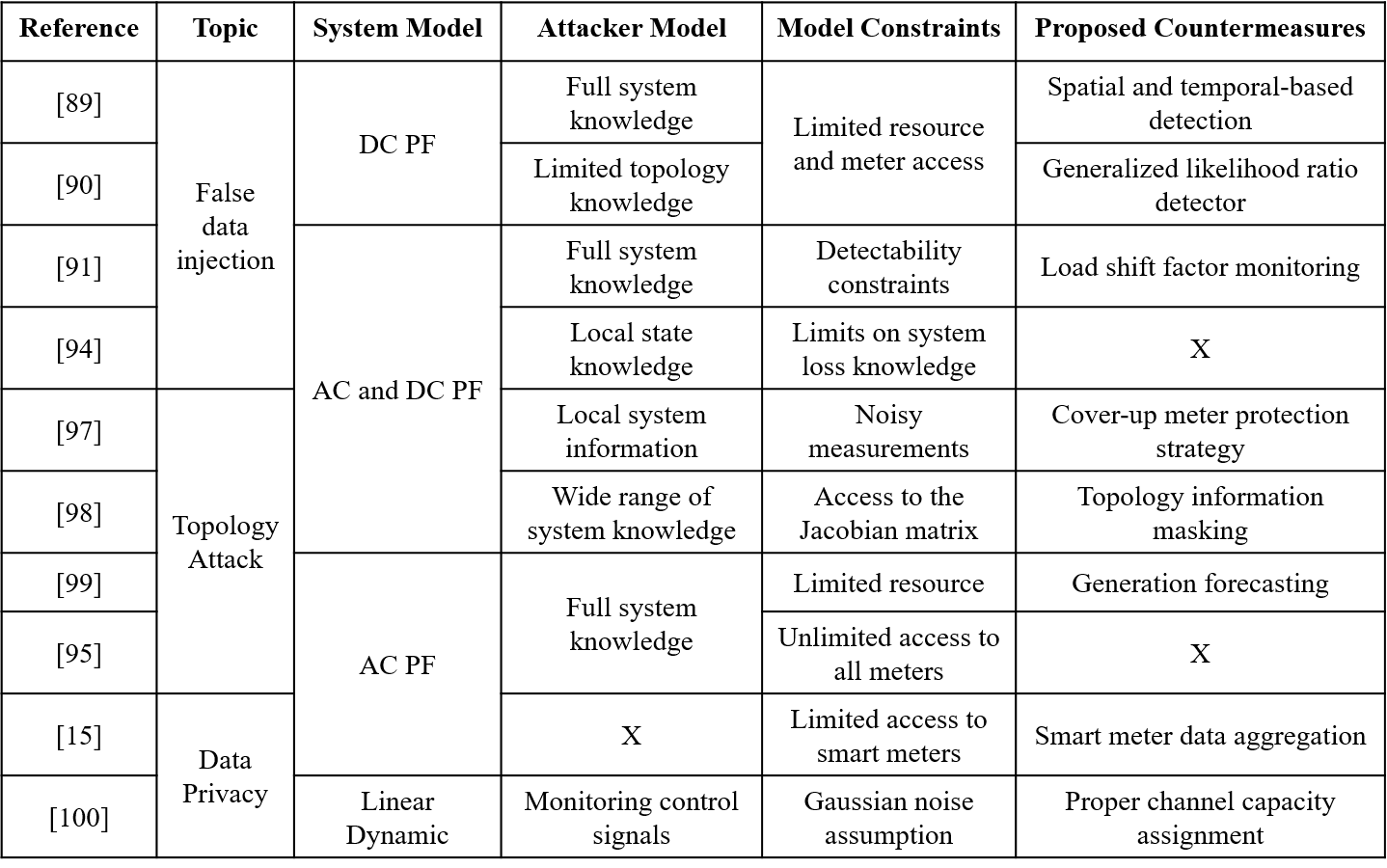}
    \label{table:CS}
\end{center}
 \vspace{-1.25em}
\end{table}
\section{Cyber-Security}\label{cs}
The vulnerability of the power system against cyber-attacks has been observed in practice. Different types of cyber-attack related to SE have been modeled and investigated in the literature: false data injection, topology attacks, and eavesdropping. In a false data injection situation, an attacker, with various degrees of knowledge on system parameters and states, alters the metered data of certain metering devices \cite{Yang2014} \cite{Li2015} \cite{Liang2016} \cite{Yu2015} \cite{Chakhchoukh2016} \cite{Deng}. In a topology attack, the attackers tend to maliciously modify the topology model data of the system \cite{Liang2017} \cite{Liu2017} \cite{Kim2013} \cite{Chakhchoukh2015} \cite{Zhang2016}. Eavesdropping defines a situation in which an unauthorized party seeks to gather system data by tapping into the communication infrastructure, compromising data privacy and confidentiality of users \cite{Li2011} \cite{Chen2016}. A classification of different papers with respect to the issue of cyber-security can be seen in Table \ref{table:CS}. It can be concluded that protecting the vital automation and monitoring systems against cyber intrusion and cyber attacks requires a holistic approach to preserve the integrity, availability, and confidentiality of DSSE at all times. Different components of an effective solution include: adversary identification (in terms of knowledge and resource levels), vulnerability assessment (critical meters, communication system integration, sensitivity of DSSE to bad data), and personnel training.

\section{Conclusion}\label{conclusion}
In this paper, we have presented an overview of the critical aspects of DSSE. Active research subjects, such as DSSE problem formulation, pseudo-measurement generation, network topology, data meter placement, renewable resource integration, and cyber-security are reviewed. Based on the survey, most recent works are more concentrated on using data-driven and machine-learning-based modifications in the conventional DSSE (for improving the accuracy, robustness, and system observability), which is a reasonable direction given the steep increase in the rate of installation of smart meters and micro-PMUs at the distribution level. Probabilistic modeling (in a data-driven context) has also attracted substantial research works, due to its capability for capturing the effects of stochastic and variable renewable resources on active distribution systems in general (and on DSSE in particular.) 

\section{Future Research Directions}\label{future}
It would be of interest to study how Demand Response (DR) programs \cite{Lu} could impact the DSSE (in terms of uncertainty and variability of customer behavior and pseudo-load generation) by incorporating retail market signals into the DSSE problem formulation. In general, integrating the price-sensitivity of active distribution networks into the DSSE becomes a valid research problem in future distribution systems with deep penetration of renewable and DR resources. In a related context, optimal power management and decision making under limited distribution system observability appears to be a largely unexplored direction for research, specially in presence of emergent technologies, such as energy storage systems and networked microgrids \cite{zwang1}\cite{zwang2}\cite{luo}\cite{luo2}. Another very recent area of interest is topology learning. Future research is needed to discover if and how topology discovery can be performed after extreme weather events \cite{Ma} as the number of data meters decreases due to communication and device failure, and the observability of the distribution system is compromised. Employing data-driven methods under extreme weather events at different stages (pre-event, during the event, and post-event) for developing system monitoring and learning techniques is another possible research direction. Thus, it would be of interest to investigate the impact of extreme events on distribution system observability and design potential solution strategies to enable effective system restoration strategies that depend on operator's real-time knowledge of system states.

\section*{Acknowledgment}
The authors are very grateful to the support from the Advanced Grid Modeling Program at the U.S. Department of Energy Office of Electricity under the grant DE-OE0000875.


\ifCLASSOPTIONcaptionsoff
  \newpage
\fi



\bibliographystyle{IEEEtran}
\bibliography{IEEEabrv,./bibtex/bib/IEEEexample}
%




%
\begin{IEEEbiography}[{\includegraphics[width=1in,height=1.25in,clip,keepaspectratio]{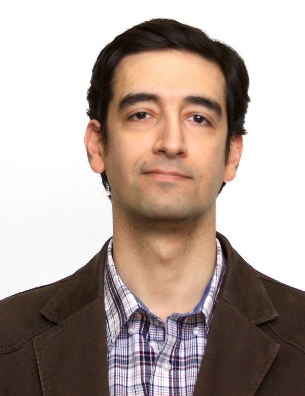}}]{Kaveh Dehghanpour}(S'14--M'17) received his B.Sc. and M.S. from University of Tehran in electrical and computer engineering, in 2011 and 2013, respectively. He received his Ph.D. in electrical engineering from Montana State University in 2017. He is currently a postdoctoral research associate at Iowa State University. His research interests include application of machine learning and data-driven techniques in power system monitoring and control.
\end{IEEEbiography}

\begin{IEEEbiography}[{\includegraphics[width=1in,height=1.25in,clip,keepaspectratio]{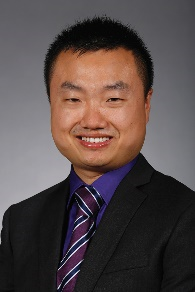}}]{Zhaoyu Wang}(S'13--M'15) is the Harpole-Pentair Assistant Professor with Iowa State University. He received the B.S. and M.S. degrees in electrical engineering from Shanghai Jiaotong University in 2009 and 2012, respectively, and the M.S. and Ph.D. degrees in electrical and computer engineering from Georgia Institute of Technology in 2012 and 2015, respectively. He was a Research Aid at Argonne National Laboratory in 2013 and an Electrical Engineer Intern at Corning Inc. in 2014. His research interests include power distribution systems, microgrids, renewable integration, power system resilience, and power system modeling. He is the Principal Investigator for a multitude of projects focused on these topics and funded by the National Science Foundation, the Department of Energy, National Laboratories, PSERC, and Iowa Energy Center. Dr. Wang received the IEEE PES General Meeting Best Paper Award in 2017 and the IEEE Industrial Application Society Prize Paper Award in 2016. Dr. Wang is the Secretary of IEEE Power and Energy Society Award Subcommittee. He is an editor of IEEE Transactions on Smart Grid and IEEE PES Letters.
\end{IEEEbiography}

\begin{IEEEbiography}[{\includegraphics[width=1in,height=1.25in,clip,keepaspectratio]{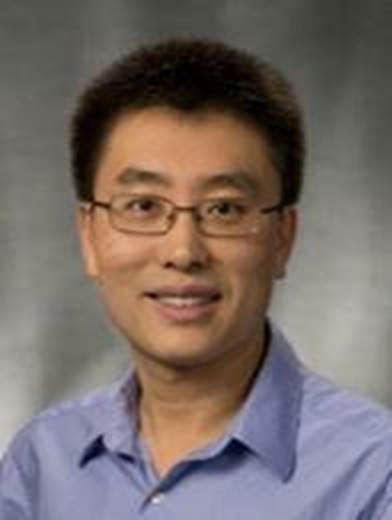}}]{Jianhui Wang}(M'07--SM'12) received the Ph.D. degree in electrical engineering from Illinois Institute of Technology, Chicago, Illinois, USA, in 2007. Presently, he is an Associate Professor with the Department of Electrical Engineering at Southern Methodist University, Dallas, Texas, USA. Prior to joining SMU, Dr. Wang had an eleven-year stint at Argonne National Laboratory with the last appointment as Section Lead – Advanced Grid Modeling. Dr. Wang is the secretary of the IEEE Power \& Energy Society (PES) Power System Operations, Planning \& Economics Committee. He has held visiting positions in Europe, Australia and Hong Kong including a VELUX Visiting Professorship at the Technical University of Denmark (DTU). Dr. Wang is the Editor-in-Chief of the IEEE Transactions on Smart Grid and an IEEE PES Distinguished Lecturer. He is also the recipient of the IEEE PES Power System Operation Committee Prize Paper Award in 2015.
\end{IEEEbiography}

\begin{IEEEbiography}[{\includegraphics[width=1in,height=1.25in,clip,keepaspectratio]{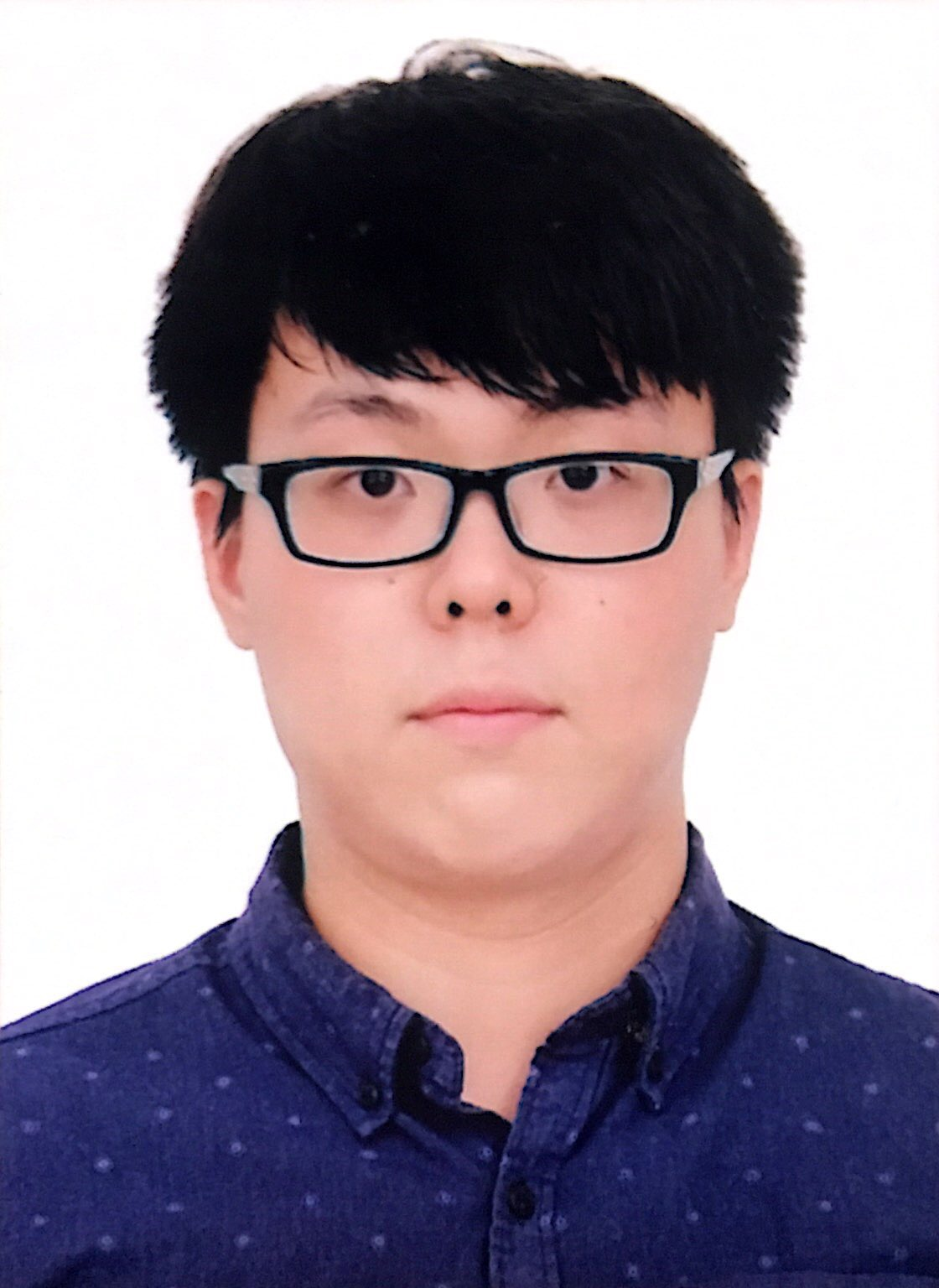}}]{Yuxuan Yuan}(S'18) received the B.S. degree in Electrical \& Computer Engineering from Iowa State University, Ames, IA, in 2017. He is currently pursuing the Ph.D. degree at Iowa State University. His research interests include distribution system state estimation, synthetic networks, data analytics, and machine learning.
\end{IEEEbiography}

\begin{IEEEbiography}[{\includegraphics[width=1in,height=1.25in,clip,keepaspectratio]{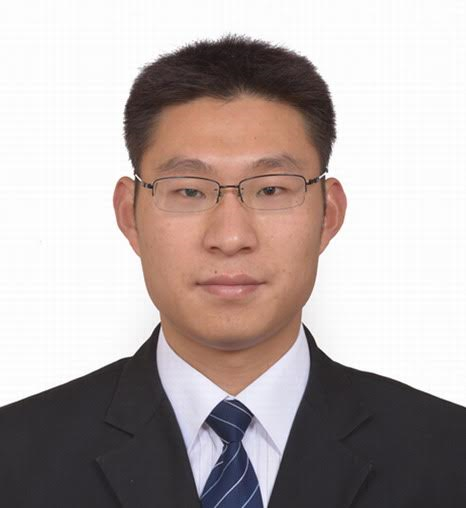}}]{Fankun Bu}(S'18) is currently pursuing his Ph.D. in the Department of Electrical and Computer Engineering, Iowa State University, Ames, IA. He received the B.S. and M.S. degrees from North China Electric Power University, Baoding, China, in 2008 and 2013, respectively. From 2008 to 2010, he worked as a commissioning engineer for NARI Technology Co., Ltd., Nanjing, China. From 2013 to 2017, he worked as an electrical engineer for State Grid Corporation of China at Jiangsu, Nanjing, China. His research interests include load modeling, load forecasting, distribution system estimation, machine learning and power system relaying.
\end{IEEEbiography}
\vfill

\end{document}